\documentclass[sigconf,screen, nonacm]{acmart}
\usepackage{balance}
\AtBeginDocument{%
  }

\usepackage{xspace}

\usepackage{epigraph}

\newenvironment{MYEQT}
  {%
    \begingroup
    \setlength{\abovedisplayskip}{4pt}%
    \setlength{\belowdisplayskip}{4pt}%
    \setlength{\abovedisplayshortskip}{2pt}%
    \setlength{\belowdisplayshortskip}{2pt}%
  }
  {%
    \endgroup
  }

\usepackage[dvipsnames]{xcolor}

 \usepackage{pifont}
 \newcommand{\cmark}{{\color{ForestGreen}\ding{51}}}%
 \newcommand{\xmark}{{\color{Maroon}\ding{55}}}%

\newcommand{\ie}{\textit{i.e.},\xspace}
\newcommand{\eg}{\textit{e.g.},\xspace}

\usepackage{amsmath}
\usepackage[caption=false]{subfig}
\usepackage{float}
\usepackage{wasysym}
\usepackage{tabularx,colortbl}

\usepackage{tikz}
\usepackage{amsmath}
\usepackage{nicefrac}
\usepackage{siunitx}
\usepackage{soul}
\usepackage{array,framed}
\usepackage{booktabs}
\usepackage[normalem]{ulem}
\usepackage{algorithm}
\usepackage{algpseudocode}
\usepackage{paralist}
\usepackage{
  color,
  float,
  epsfig,
  wrapfig,
  graphics,
  graphicx,
}

\usepackage{setspace}
\usepackage{enumerate}
\usepackage{xparse} 
\usepackage{multirow}
\usepackage{csvsimple}
\usepackage{balance}
\usepackage{hyperref}
\usepackage{cleveref}
\usepackage{listings}

\usepackage{enumitem}

\crefformat{section}{\S#2#1#3}
\crefname{figure}{Figure}{Figures}
\crefname{appendix}{Appendix}{Appendices}
\crefname{table}{Table}{Tables}
\crefname{table2}{Table}{Tables}
\crefname{algorithm}{Algorithm}{Algorithms}
\crefname{listing}{Listing}{Listings}
\crefname{theorem}{Theorem}{Theorems}
\crefname{thm}{Theorem}{Theorems}
\crefname{lemma}{Lemma}{Lemmata}
\crefname{equation}{Eqt.}{Eqts.}
\crefformat{Grammar}{Grammar #1}

\usepackage{
  tikz,
  pgfplots,
  pgfplotstable
}

\usetikzlibrary{
  shapes.geometric,
  arrows,
  external,
  pgfplots.groupplots,
  matrix
}

\pgfplotsset{compat=1.9}

\usepackage{mathtools,}

\definecolor{keywordcolor}{rgb}{0, 0.3, 0.7}
\definecolor{commentcolor}{rgb}{0.5, 0.5, 0.5}
\definecolor{stringcolor}{rgb}{0.58, 0, 0.82}
\newcommand{\sysname}{DiVerify\xspace}

\newcommand{\myparagraph}[1]{\vspace{0.29cm}\textbf{#1:}\xspace}

\newif\ifARXIV
\ARXIVtrue
\newif\ifDEBUG
\DEBUGfalse

\ifDEBUG
    \newcommand{\CO}[1]{{\color{brown} Chinenye says: #1}}
    \newcommand{\JD}[1]{{\color{blue} Jamie says: #1}}
    \newcommand{\santiago}[1]{{\color{red} Santiago says: #1}}
\else
    \newcommand{\CO}[1]{}
    \newcommand{\JD}[1]{}
    \newcommand{\santiago}[1]{}
\fi

\ifARXIV

\else

\setcopyright{cc}
\setcctype{by}
\acmDOI{10.1145/3832783.3834333}
\acmYear{2026}
\copyrightyear{2026}
\acmISBN{979-8-4007-2882-2/2026/10}
\acmConference[ASE '26]{Proceedings of the 41st IEEE/ACM International Conference on Automated Software Engineering}{October 12--16, 2026}{Munich, Germany}
\acmBooktitle{Proceedings of the 41st IEEE/ACM International Conference on Automated Software Engineering (ASE '26), October 12--16, 2026, Munich, Germany}
\acmSubmissionID{ase26main-p30-p}
\received{2026-03-27}
\received[accepted]{2026-06-18}
\fi
\begin{document}


\title[Context-Aware Trust Verification for Identity-Based Software Signing]{Context-Aware Trust Verification \\ for Identity-Based Software Signing}

\author{Chinenye Okafor}
\affiliation{%
  \institution{Purdue University}
  \city{West Lafayette}
  \state{IN}
  \country{USA}}
\email{okafor1@purdue.edu}

\author{James C. Davis}
\affiliation{%
  \institution{Purdue University}
  \city{West Lafayette}
  \state{IN}
  \country{USA}}
\email{davisjam@purdue.edu}

\author{Santiago Torres-Arias}
\affiliation{%
  \institution{Purdue University}
  \city{West Lafayette}
  \state{IN}
  \country{USA}}
\email{santiagotorres@purdue.edu}

\renewcommand{\shortauthors}{Chinenye Okafor, James C. Davis \&  Santiago Torres-Arias}
\begin{abstract}
Modern software release infrastructure uses identity-based software signing to associate software artifacts with a known identity.
For example, registries such as npm and Docker Hub rely on Sigstore's identity-based software signing in their artifact provenance workflows.
However, existing verification procedures only provide evidence of a signer's identity, but not the conditions under which signing occurred.
As a result, compromised credentials, identity providers, or signing tools can still produce signatures that pass verification.
This gap prevents software engineering tools (\eg package registries, dependency analysis tools, deployment systems, and CI/CD verification stages) from enforcing verifiable context-aware trust policies for software signing.

We present \sysname, a framework for automated verification of software signing conditions.
A \sysname signature binds identity claims from multiple independent scope providers, signing-environment attestation from a TEE-isolated daemon, and the signing key into a single verifiable object.
A machine-checkable policy language specifies the identity, authentication, and environment conditions
required for accepting a signature.
We formally analyze DiVerify and show that it is sound, prototype it across three deployment modes, and demonstrate overhead under 400ms, with storage overhead of less than 1.6\% of average PyPI package size.
Case study integrations with existing systems confirm DiVerify composes well. 


\end{abstract}


\maketitle


\section{Introduction}

Modern software applications rely on third-party artifacts obtained through a software supply chain~\cite{sonatype2024state}.
As a result, software supply chains have become a prominent attack surface:
  adversaries target maintainer accounts~\cite{gu2023investigating},
  identity providers~\cite{microsoft2024midnight},
  build infrastructure~\cite{fourne2023s, ohm2020backstabber},
  and release tooling~\cite{codecov2021}
  to cause malicious artifacts to appear legitimate.
These risks have made software artifact verification a central software engineering concern~\cite{kalu2025industry, kalu2025johnny, schorlemmer2024signing}.

Software signing is a standard mechanism for verifying software artifacts.
Current software signing approaches verify signer \textit{identity}: a producer signs an artifact, and a consumer verifies that the artifact is associated with an expected signer.
This is true both in conventional signing systems such as PGP (where identity is inferred from control of a signing key), and in modern identity-based systems~\cite{newman_sigstore_2022, keyfactor} (where ephemeral signing keys are bound to authenticated identities).
These verification procedures provide evidence only of \emph{who} signed, not \emph{under what conditions} signing occurred.
In incidents such as ESLint~\cite{eslint2018postmortem} and Codecov~\cite{codecov2021}, attackers exploited this trust gap to produce artifacts whose signatures were acceptable to verifiers, yet were generated under conditions the maintainer did not authorize.
When engineers want their release processes to enforce such contextual trust requirements, they currently rely on manual review, out-of-band trust judgments, or implicit assumptions about the signing infrastructure rather than automated, machine-checkable verification.

We present \sysname, a framework for verifiable context-aware trust policies in software signing.
\sysname augments software signing with contextual evidence that can be checked automatically during artifact verification.
Its first mechanism is a \textit{cryptographically verifiable proof} that binds identity claims from independent providers to signing-context attestation and the signing key used for the artifact.
Second, \sysname has a \textit{machine-checkable policy language} that specifies the necessary identity, authentication, and environment conditions for a signature to be accepted.
Automated verifiers check a proof against a policy before accepting an artifact.

We instantiate \sysname and evaluate it across legacy-compatible workflows to a fully attested workflow. These modes capture tradeoffs between deployability and security, allowing incremental adoption as infrastructure supports. We formally prove that DiVerify is sound under bounded provider compromise.
Across deployment modes, \sysname introduces low overhead despite the security guarantees it provides.
Signing completes in under 400 ms, with storage overhead of less than 1.6\% of the average PyPI package. We perform exploratory case studies integrating DiVerify with Sigstore, OpenPubkey, and RSTUF~\cite{rstuf}, showing that DiVerify fits into existing signing and distribution infrastructure with minimal changes.

In summary, we contribute:

\begin{itemize}[itemsep=3pt,topsep=3pt]
    \item \textit{Signing-context trust model.}
    We present a verifier-oriented model for context-aware software signing that adapts prior authentication and attestation to this setting, including \emph{independence} as a required property of signing-context evidence.

    \item \textit{\sysname framework and policy verification.}
    We design \sysname, a framework for verifiable context-aware trust policies in software signing, combining cryptographically verifiable proofs of signing context with machine-checkable policy evaluation during artifact verification.

    \item \textit{Evaluation.}
    We implement \sysname, prove that it is sound under practical assumptions, and empirically evaluate its performance and storage costs.
    We further conduct exploratory case studies suggesting feasible integration with existing signing and distribution infrastructure.
 
\end{itemize}

\noindent
\textbf{Significance:}
\sysname shifts artifact verification from checking signer identity alone to enforcing explicit, machine-checkable trust requirements about signing conditions.
This capability allows engineers to automate trust decisions that currently require implicit assumptions or manual review.
Although we evaluate \sysname in the context of identity-based signing, its broader significance lies in enabling release infrastructure to verify artifacts against expected signing conditions instead of relying on implicit trust.

\section{Background}
\label{sec:background}

This section reviews
  software signing (\cref{sec:bg-signing}),
  authentication mechanisms (\cref{auth-mechanisms}),
  and our threat model (\cref{sec:bg-threats}).

\subsection{Software Signing and Verification}
\label{sec:bg-signing}

\subsubsection{Signing Workflow}
Modern software ecosystems distribute artifacts through automated release pipelines that build, sign, and publish packages. 
These workflows involve source repositories, third-party dependencies, build infrastructure, CI/CD pipelines, packaged artifacts, and distribution registries, forming a complex software supply chain~\cite{schorlemmer2024signing, okafor_sok_2022, kumar2017security, synopsys_osra}.
Because downstream systems typically receive only the final artifact, they must assess its trustworthiness with incomplete knowledge of how it was produced.

Software signing is a standard mechanism for artifact verification.
A producer (\textit{signer}) signs an artifact, and a consumer (\textit{verifier}) uses public-key verification to check that the signature is valid for that artifact and corresponds to an expected signer~\cite{rivest1978method,kalu2025software}.
Traditional signing systems such as OpenPGP~\cite{openpgp} rely on long-lived signing keys and manual trust distribution, which creates usability challenges and risks of compromise~\cite{chandramouli2013cryptographic, unger2015sok}.
Recent systems
  such as Sigstore~\cite{newman_sigstore_2022} instead adopt \emph{identity-based software signing}, in which ephemeral signing keys are generated for individual signing events and bound to identities authenticated through standard protocols such as OpenID Connect~\cite{openidconnect-Sakimura}. OpenPubkey~\cite{heilman_OpenPubkey_2023} extends identity-based signing with cosigners but remains limited to identity binding, leaving broader signing conditions unenforceable.

\begin{figure}[h]
    \centering
    \includegraphics[width=\linewidth]{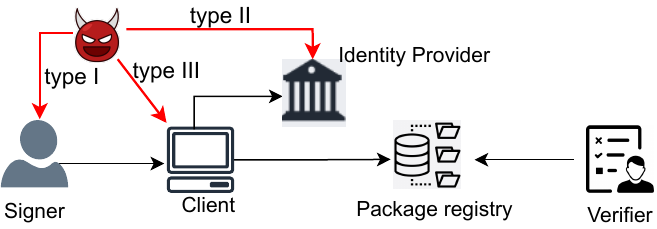}
    \caption{
    Identity-based code signing flow, together with attacks that compromise existing systems (\cref{sec:bg-threats}).}
    \label{fig:code-signing}
\end{figure}

A typical signing flow is illustrated in~\cref{fig:code-signing}.
From the producer's view:
\begin{inparaenum}
    \item First, the signer initiates a signing request through a client;
    \item Next, the client redirects the signer to an identity provider (IdP) for authentication;
    \item Then, the signer authenticates, an identity token is issued, and the token is returned to the client;
    \item Optionally, the client requests a signing certificate that binds the authenticated identity to a signing key; and
    \item Finally, the client signs the software artifact and uploads to a registry.
\end{inparaenum}

As a consumer, the verifier obtains from a source (\eg a package registry) the software artifact $SW_a$, its signature $S_{sw}$, and the corresponding public key $K_p$, which we write as the tuple $(S_{sw}, K_p, SW_a)$.
Using public-key cryptography, the verifier confirms $S_{sw}$ is valid signature on $SW_a$ under $K_p$.
The verifier also evaluates identity attributes associated with the public key $K_p$, \eg the signer's email address and the issuing IdP, against locally configured trust expectations.
These checks determine the artifact's acceptance.

For this verification task, NIST SP 800-63B identifies four properties for trusting authentication results~\cite{temoshok2025digital}. The most relevant here are \emph{replay resistance}, binding claims to the current signing session and \emph{verifier impersonation resistance}, ensuring claims genuinely originate from the expected provider rather than an attacker-controlled stand-in. The remaining two properties, MitM and verifier compromise resistance, are handled by authenticated transport and tool-integrity design respectively.

\subsubsection{Validating Signatures}

However, existing signing systems provide evidence primarily about \emph{who} signed, not \emph{under what conditions} signing occurred~\cite{newman_sigstore_2022, heilman_OpenPubkey_2023, keyfactor}.
The NIST authentication properties above help explain when a verifier can trust an authentication result, but they do not by themselves provide a verifier-oriented model for software-signing context.
NIST SP 800-204D further recognizes, in the context of CI/CD security, that trustworthy supply-chain evidence should be generated
  by components with stronger trust than the build itself, and that such evidence may be evaluated against policy~\cite{chandramouli2024strategies}.
However, that guidance omits a concrete verifier-oriented model for software-signing verification~\cite{chandramouli2024strategies}.

For software signing, this leaves an important capability gap.
A verifier may be able to confirm that a signing key was used and that the associated identity appears acceptable, yet still lack verifiable evidence about the signing context.
In particular, evidence about the signing environment must be meaningful even if the signing tool under evaluation has been compromised.
\emph{We therefore treat independence as a required property of signing-context evidence}.
Evidence about the signer, signing tool, or environment must be produced outside that tool's control if it is to remain meaningful under compromise.
Current signing systems do not provide such verifier-checkable context evidence, so downstream acceptance decisions still depend on manual review, out-of-band trust judgments, or assumptions about the signing infrastructure.

\subsection{Authentication Mechanisms}
\label{auth-mechanisms}

Authentication mechanisms are commonly grouped into three classes of claims~\cite{huang2010generic,mahadi2018survey}.
\emph{Identity claims}, such as OpenID Connect (OIDC)~\cite{noauthor_how_2023}, assert who the signer is through statements from a trusted identity provider.
\emph{Possession claims}, \eg FIDO2/WebAuthn challenge-response with a hardware private key~\cite{YubicoPythonFido2}, prove control of a protected secret.
\emph{Contextual claims}, as with TPM~\cite{tpm-me} or device fingerprinting~\cite{rizzo2021unveiling},
assert properties of the execution environment.

Software signing systems typically consume only identity evidence.
This supports accountability, but not verification of the conditions under which signing occurred.
Although authentication mechanisms can also provide possession and contextual evidence, current signing verification procedures generally do not expose those signals, leaving some trust requirements unenforceable.

\subsection{Attacks and Threat Model}
\label{sec:bg-threats}

The capability gap above is consequential because the software-signing ecosystem is itself an attack surface.
In incidents such as
  ESLint~\cite{eslint2018postmortem},
  Midnight Blizzard~\cite{microsoft2024midnight},
  and
  Codecov~\cite{codecov2021},
  attackers were able to produce artifacts whose signatures remained acceptable to existing verifiers, yet were generated under conditions the maintainer did not authorize.

We consider attackers to be successful if they can cause a verifier to accept a malicious artifact $(SW_a)$.
We consider three attack types reflecting different ways an adversary can obtain signing authority:

\begin{enumerate}[leftmargin=2em,itemsep=2pt,topsep=2pt]
    \item \textit{Type I: Credentials.}
    The attacker compromises a signer's account, \eg through phishing or session hijacking, and then authenticates and signs on the signer's behalf, as in ESLint~\cite{eslint2018postmortem}.

    \item \textit{Type II: Identity Provider.}
    The IdP is compromised or malicious and can issue identity tokens for a signer who did not authenticate, allowing attackers to impersonate signers without interacting with the legitimate user, as in Midnight Blizzard~\cite{microsoft2024midnight}.

    \item \textit{Type III: Signing Tool.}
    The attacker tampers with the client software so that it signs software beyond what the signer originally intended, allowing unauthorized signatures to be produced under apparently valid identities, as in Codecov~\cite{codecov2021}.
\end{enumerate}

\noindent
These three attack types follow from~\cref{fig:code-signing}.

This taxonomy reflects a stronger threat model than that used in prior software supply-chain literature.
Prior work has modeled threats to particular segments of the pipeline, including source-code integrity~\cite{vu2021lastpymile, gonzalez_anomalicious_2021}, build-system compromise~\cite{fourne2023s, ohm2020backstabber}, maintainer account compromise~\cite{zimmermann2019small, samuel2010survivable, gu2023investigating, huang2024donapi}, package registries~\cite{Trishank2016, zimmermann2019small, zahan_what_2022, wu2023understanding, nikitin2017chainiac}, mirror compromise~\cite{cappos_look_2008}, package distribution~\cite{gao2024pyradar}, and dependency-resolution attacks~\cite{neupane2023beyond, vu2020typosquatting}.
However, many existing models assume that the identity provider and signing client are trusted, and rely on transparency logs or related mechanisms to detect misuse after a malicious signing event has already occurred~\cite{al-bassam_contour_2018, blauzvern2023nowhere, laurie2014certificate}.
In light of the real attacks described above, that is no longer sufficient for downstream verifiers, which must decide whether to trust an artifact in real time.
Components traditionally treated as trusted are now realistic targets.
We therefore adopt a \textit{stronger threat model} that treats the identity and signing ecosystem itself as an attack surface.
Our goal is to enable verifiers to distinguish between signatures produced under expected signing conditions and signatures produced under policy-violating conditions, even when the resulting artifact and signer identity appear acceptable under existing verification procedures.

\section{DiVerify Overview}
\label{sec:design}

DiVerify seeks to enforce resilience against credential compromise (S1), identity-provider compromise (S2), and compromised signing clients (S3) such that no single compromised signing component should be sufficient to produce an artifact that passes verification.
To achieve this, it provides the notion of \emph{context-aware signing}. 



\begin{figure}
    \centering
    \includegraphics[width=\linewidth]{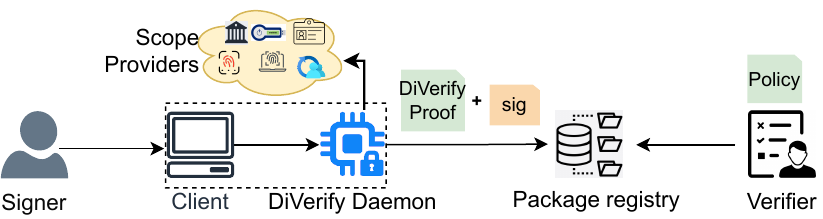}
    \caption{
    DiVerify overview. Scope providers assert signer claims that, together with client integrity attestation, form DiVerify proof. Verifiers validate the proof against a policy to decide whether to accept the signed artifact. 
    }
    \label{fig:diverify_architecture}
\end{figure}

\subsection{Context-Aware Signing}
\label{sec:trustmodel}


Context-aware signing is an extension to identity-based signing systems, to include multiple trust signals (or scopes) from various sources.
This mechanism follows long-established secure design methodologies. 
For example, in authentication, practitioners achieve resilience to credential compromise through multi-factor authentication (MFA), which establishes identity with a provider using multiple factors. 
However, in such a setting, verifiers have no way to confirm that these factors were actually presented and must trust the identity provider to report authentication outcomes honestly. 
As a result, a compromised provider can still issue valid tokens regardless of the factors used.

DiVerify extends this principle for software signing in three ways: first, across independent providers so that no single provider compromise suffices; and second, beyond identity to the signing environment itself; and third, it enables downstream verifiers to transparently verify these factors. 

The resulting scheme achieves our desired security properties and underlies three requirements:

\begin{enumerate}[leftmargin=2em,itemsep=2pt,topsep=2pt]
    \item \textbf{Multi-scope identity.} Signer identity must be corroborated by claims from multiple independent providers, such that compromise of any single provider is insufficient to produce an accepted signature.

    \item \textbf{Environment integrity.} The signing environment must be attested by a component isolated from the signing client, providing independent evidence that the signing tool was unmodified at signing time.

    \item \textbf{Policy Governance.} Artifact acceptance must be conditioned on explicit, machine-checkable policy specifying the identity and contextual information to trust a signature.
\end{enumerate}

While the high-level idea is intuitive, realizing it in a way that is secure, composable, and verifiable is non-trivial. The remainder of this section develops the design needed to achieve this.

\subsection{System Model}\label{sys-model}
We begin with the system components and adversary assumptions shaping our design.
DiVerify consists of
  a \textit{signer},
  a signing \textit{client},
  \textit{scope providers} that issue claims,
  an \textit{attester} that reports the client's integrity,
  a \textit{package registry} that hosts artifacts and proofs,
  and
  a \textit{verifier} that holds a policy and evaluates proofs against it (\cref{fig:diverify_architecture}).

The adversary controls a subset of the signer's credentials and scope providers, as defined in~\cref{sec:bg-threats}. The adversary aims to produce a software artifact and accompanying proof that validates against the expected policy while violating the intended signing conditions.

We assume that
  verifiers execute the verification procedure correctly;
  the attester operates outside the signing client's trust boundary and functions correctly;
  and
  not all independent scope providers are simultaneously compromised.
We do not consider failures of underlying cryptographic primitives.

\subsection{Composing Multiple Providers}
\label{sec:multiscope}

A central element of the DiVerify model is the ability to compose trust signals. 
A natural approach to realizing the first requirement in~\cref{sec:trustmodel} is to require evidence from multiple independent providers. 
Intuitively, combining identity providers (e.g., OIDC), possession proofs (e.g., hardware keys), and contextual signals (e.g., device fingerprints) should improve security. 
However, naively aggregating such evidence is insufficient: without secure composition, attackers can mix and replay claims across sessions or impersonate providers.

We identify three elements for provider composition:

\myparagraph{1) Provider independence} Compromise of one provider should not allow impersonation of others. 
In practice, providers may share underlying trust roots (e.g., accounts linked to the same email), in which case requiring multiple such providers does not provide independent security guarantees. Independence is therefore a property of the set of providers used in a policy, rather than of any provider.

\myparagraph{2) Session binding} All evidence must be cryptographically bound to a single signing session. Otherwise, an attacker may combine stale or replayed claims from different sessions. For example, an attacker could replay a previously captured OIDC claim while supplying a fresh hardware-backed proof, thereby satisfying multiple requirements without controlling all factors simultaneously.

To prevent such attacks,  we force each provider to produce evidence that is bound to a verifier-supplied session nonce. This ensures that all claims correspond to the same logical signing event.

\myparagraph{3) Claim authenticity} The verifier must be able to determine that each claim originates from the claimed provider. 
Otherwise, an attacker may fabricate or substitute claims.
This requirement can be satisfied in two distinct ways.
First, native provider authenticity, where the provider directly signs its output (\eg OIDC tokens, hardware attestations). 
Second, a trusted entity (\ie a mediator within a TEE) can collect and attest to provider outputs instead. 

To satisfy session binding and claim authenticity at the level of individual providers, a provider must support (i) binding its output to a session nonce and (ii) producing evidence that can be attributed to its source. These two properties define a minimal model of providers in our setting. Independence, by contrast, is enforced at the policy level across providers. This yields four categories of providers as shown in~\cref{tab:providers}.

\begin{table}[h]
\centering
\caption{
Scope provider taxonomy derived from \S\ref{auth-mechanisms}.
}
\label{tab:providers}
\begin{tabular}{lcc}
\toprule
\textbf{Category} & \textbf{Replay Resistant} & 
\textbf{Authenticable} \\
\midrule
Verifiable (VP)  & \cmark & \cmark \\
Static (SP)      & \xmark & \cmark \\
Opaque (OP)      & \cmark & \xmark \\
Untrusted (UT)   & \xmark & \xmark \\
\bottomrule
\end{tabular}
\end{table}

Providers that are not both session-bindable and authenticable require additional mechanisms.
\emph{Static providers }(SPs) produce fixed outputs (\eg device fingerprints) that cannot directly incorporate a nonce.
To enable session binding, their outputs must be combined with the session nonce and authenticated by a trusted component.
As static providers cannot incorporate a nonce, an additional mechanism is required to bind their outputs to the session.
This necessitates a trusted mediator that authenticates the commitment by signing the hash of the outcome given a nonce, or a zero-knowledge proof demonstrating knowledge of a static value bound to the session.
\emph{Opaque providers} (OPs) produce no fresh outputs nor native authenticity.
These can be incorporated by collecting their outputs inside a TEE, which attests to the measurement process and binds the result to the session.
In both cases, the verifier relies on the trustworthiness of the mediator or TEE, not on the provider itself.

\subsection{Tool Integrity Attestation}
\label{sec:attestation}

Interacting with providers must be carried out by a trusted party, which will also measure and provide proof that the signing context is trusted.
This achieves requirement two from~\cref{sec:trustmodel}. 
Under a Type III adversary, a compromised tool can satisfy all identity and session requirements while signing unintended artifacts. 
Closing this gap requires evidence about the tool integrity.

As a strawman, consider having the signing tool report its own identity, for example by including a hash of its binary with the artifact signature.
This is unsound: any claim produced inside the signer-controlled environment can be forged by a compromised tool, including claims about its own identity or execution state. Thus, tool integrity must be measured by an entity that is structurally outside the signing environment’s control.

DiVerify therefore requires \textit{independently verifiable execution evidence as a condition of signature validity}.
A DiVerify signature is accompanied by an attestation from an external trust anchor with evidence about the execution context in which signing occurred.
Verification succeeds only if this attestation verifies under a verifier-accepted trust root and satisfies policy.

The strength of this evidence depends on the trust boundary of the attesting mechanism.
Different deployments admit different points in a trust-boundary/deployability tradeoff:
  some rely on organizationally-enforced measurement,
  some on hardware roots of trust,
  and
  some on hardware-isolated execution.
DiVerify treats the evidence source abstractly as an independent attestor and defines a uniform verification interface over any attestation mechanism satisfying the required independence property.

Three representative instantiations are shown in~\cref{tab:attestation}.

\begin{table}[h]
\centering
\caption{Representative environment-measurement mechanisms and their trust-boundary/deployability tradeoff.}
\label{tab:attestation}
\small
\begin{tabular}{llc}
\toprule
\textbf{Mechanism} & \textbf{Trust Boundary} & \textbf{Deployability} \\
\midrule
Externally-enforced hash & Org.\ governance      & Universal \\
TPM measured boot                    & Hardware (boot chain) & Requires TPM \\
TEE attestation                      & Hardware isolation    & Requires TEE \\
\bottomrule
\end{tabular}
\end{table}

At the weakest end, an independent process measures the cryptographic hash of the signing tool.
Here, the integrity of the measurement depends on organizational governance correctly enforcing the measurement process.
This scheme is easily deployed, but a Type III attacker may tamper with the measurement process or substitute the tool between measurement and execution.

A stronger construction uses TPM-based measured boot, which records system state in hardware before software executes.
This roots the measurement in the hardware and boot chain rather than in organizational process.
Because the measurement is established prior to execution, it cannot be modified by software running afterward, including a compromised signing tool~\cite{tpm-me, microsoft_windows11_system_requirements, schear2016bootstrapping}.

At the strongest end, TEE attestation binds measurement and execution together inside a hardware-isolated environment.
The attestation is produced by a manufacturer-rooted key that cannot be forged by a Type III attacker controlling the client environment.
DiVerify requires two properties from such TEEs: \textit{remote attestation and custom report-data binding},
as supported by architectures such as Intel SGX~\cite{costan2016intel} and ARM TrustZone~\cite{ngabonziza2016trustzone}.

\subsection{Exposing Signing Context to Verifiers}
\label{sec:proof}

Once evidence is collected and client tooling is measured, \sysname can provide a bundle exposing the signing context for verification.
That is, a \emph{DiVerify proof} bundles this evidence into a single cryptographic object bound to a specific 
signing event:
\begin{MYEQT}
\[
dvp = \langle scopes, attestation, pk \rangle
\]
\end{MYEQT}
where $scopes$ is the set of authenticated scope claims collected during the signing session, $attestation$ is the environment measurement along with its trust boundary evidence, and $pk$ is the ephemeral signing key used to produce the artifact signature. 
The proof is cryptographically bound to the signing session, preventing evidence from one session being detached and reused in another. The proof is also distributed alongside the artifact signature, giving verifiers direct, independently checkable evidence about the conditions under which signing occurred, rather than requiring them to implicitly trust the surrounding signing infrastructure. A DiVerify policy (\S\ref{sec:policy-semantics-and-lifecycle}) then specifies what this evidence must contain for a signature to be accepted.

\subsection{Policy Model and Lifecycle}\label{sec:policy-semantics-and-lifecycle}
To achieve Policy Governance, verifiers must match proofs against a policy that describes the signals expected and their values. Because these trust requirements may change over time, the policy model must also support authenticated updates, freshness, and revocation.


\subsubsection{DiVerify Policy Language}
\label{sec:policy-lang}
Existing signing systems lack a general policy framework for expressing verifier trust requirements over signing conditions. Although related domains (\eg Rego, XACML) offer highly expressive policy languages, they present usability challenges~\cite{opdebeeck2026empirical, rezvani2019analyzing} due to their complexity and steep learning curve. We therefore adopt a minimal set of constraints, following approaches validated by prior work~\cite{torres-arias_-toto_2019}, which has achieved wide adoption in the wild. 

Trust requirements vary across deployments. For example, a maintainer signing on a personal device may require both identity and possession claims, together with some contextual (\eg device fingerprint) evidence that signing occurred on a specific trusted machine. In a remote environment, the same maintainer may still require identity and possession claims, but not device-specific evidence. The policy language therefore must express both strict requirements and alternative trust paths when some contextual evidence is unavailable or changes.

The policy language is designed to be: (1) structurally aligned with the DiVerify proof, so that evidence can be evaluated unambiguously; (2) expressive enough to capture realistic signing workflows, including alternative acceptable trust paths; and (3) safe, in the sense that a policy cannot be satisfied by mixing claims from different signers or inconsistent signing contexts.

Accordingly, a policy consists of three parts:
\begin{MYEQT}
\[
P = \langle Scope, Attest, Rule \rangle
\]
\end{MYEQT}
where \emph{Scope} specifies claim types and values that a verifier should accept for a signer. For each signer $u$, it maps claim type $t$ to a set of acceptable values. At verification time, the predicate \(\textsf{hasScope}(t, v) \) holds if the proof contains a valid authenticated claim of type $t$ with value $v$.

$Attest$ specifies acceptable signing environment evidence:
\begin{MYEQT}
\[
Attest = \langle M, B, Q \rangle
\]
\end{MYEQT}
where $M$ is the expected environment measurement, $B$ binds the attested environment to the signing key and proof context, and $Q$ specifies the attestation verification procedure.
The predicate \(\textsf{validAttest}\) holds if the attestation verifies under $Q$, reports measurement $M$, and has the expected binding value $B$.

The $Rule$ is Boolean expression over claim predicates:
\begin{MYEQT}
\[
\begin{aligned} Rule =~ & \textsf{match}(c,t,v) \mid Rule_1 \land Rule_2 \mid Rule_3 \lor Rule_4 \mid \lnot Rule_5 \end{aligned}
\]
\end{MYEQT}

where \(\textsf{match}(c,t,v)\) holds when signer $u$ presents a claim $c$ of type $t$ with value $v$. A rule is satisfied only if the entire expression evaluates to true for a signer $u$; thus evidence from different signers cannot be combined to satisfy one policy.

Together $Scope$, $Attest$, and $Rule$ let maintainers specify both the required evidence and the acceptable trust paths, while preserving single signer consistency. We formalize the corresponding verification semantics in~\cref{formal-model}.

\subsubsection{Policy Metadata, Update, and Revocation}\label{policy-update-and-revocation}
In addition to policy semantics, verifiers must determine whether a policy is authentic, fresh, and not superseded. We therefore associate each policy with a signed metadata that supports authenticated updates, freshness checking, rollback protection, and revocation.

A policy instance is a pair $(Meta, P)$ where P is the policy above and Meta is a signed record:
\begin{MYEQT}
\[
Meta = \langle \textsf{version}, \textsf{epoch}, \textsf{issuedAt}, \textsf{expiresAt}, h_P, sig_{pol}, sig_{root} \rangle
\]
\end{MYEQT}

\begin{figure}
    \centering
    \includegraphics[width=0.85\linewidth]{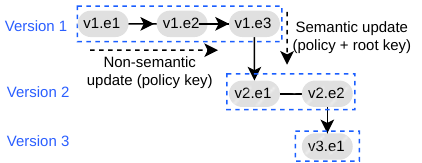}
    \caption{Policy metadata evolution. \textit{Vertical changes}: version updates that alter policy semantics. \textit{Horizontal changes}: epoch updates that only refresh metadata.}
    \label{fig:policy-evolution}
\end{figure}

Here, $h_P$ binds the metadata to a specific policy.  $version$ captures semantic updates to the policy (as shown in~\cref{fig:policy-evolution}), and $epoch$ captures the non-semantic metadata refreshes. 

Any changes that affect the policy meaning, for example, changing trusted signer, accepted claims, or Rule structure increments $version$. Because such updates change the trust configuration, a new version must be authorized by the policy key and a higher-privileged root key. This dual signature requirement ensures that no single actor can unilaterally introduce or weaken trust conditions.

\begin{figure*}
    \centering
    \subfloat[Legacy-Compatible Mode (no client attestation).\label{fig:legacy_mode}]{
        \includegraphics[width=.45\linewidth]{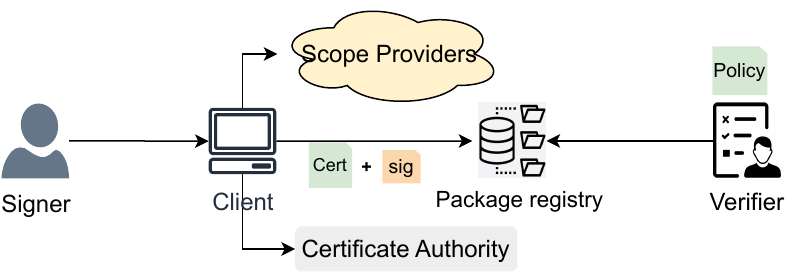}
    }
    \hfill
    \subfloat[Trusted Authentication Mode, protecting against compromised clients. \label{fig:trusted_auth_mode}]{
        \includegraphics[ width=.45\linewidth]{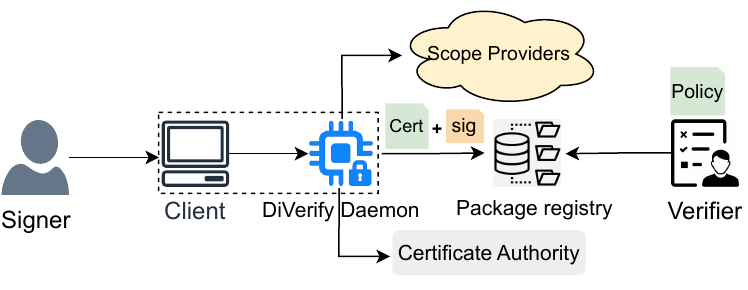}
    }
    \caption{
    Two integration modes for using DiVerify with existing signing systems, in addition to the Core Mode (\cref{fig:diverify_architecture}).
    }
\end{figure*}

Operational updates do not change policy meaning and only extend validity, so they increment $epoch$ while leaving the policy and the $version$ unchanged (as shown in~\cref{fig:policy-evolution}). Because these updates do not alter trust semantics, they require only the current policy key.
We treat policy key rotation as a semantic update since it changes the authority allowed to issue future metadata; it then requires root key authorization.

The metadata timestamps $issuedAt$ and $expiresAt$ provide freshness guarantees and limits the replay of stale metadata. A verifier accepts policy metadata only when the current time lies within its validity interval. To prevent rollback, verifier stores the highest previously accepted $(version, epoch)$ pair for each policy and rejects any metadata with a lower value.
This small amount of persistent state suffices to ensure strong rollback resistance in offline and adversarial settings, while imposing negligible storage overhead, which is less state than the manifest and metadata caches routinely maintained by software verification tools, such as APT~\cite{hertzog2014debian}, DNF~\cite {DNF:command-ref}, and TUF~\cite {gittuf_repo} clients.
In deployments where such systems already manage trusted metadata, DiVerify's policy metadata can be integrated or subsumed into their existing mechanisms for distribution and caching (\cref{sec:discussion-next}).

Revocation is expressed through a new semantic version that removes or tightens previously trusted conditions. Because verifiers enforce monotonic version advancement, superseded policies cannot be reinstated by replaying older metadata.

\section{Implementation}

We prototype our core \sysname architecture and two variants to support \sysname's integration into existing systems.

\subsection{Deployment Modes}\label{deploy-mode}
To illustrate how DiVerify integrates with existing signing systems under existing trust architectures, we implement three deployment modes that differ in where trust decisions are enforced.
These modes correspond to different integration points within existing signing infrastructures rather than distinct systems.

\myparagraph{Legacy-Compatible mode (\cref{fig:legacy_mode})}
\sysname adds multi-scope identity verification but does not enforce signer runtime attestation to easily integrate into identity-based software signing systems such as Sigstore. 
In this mode, Sigstore is integrated end-to-end in its standard path: OIDC identity, Fulcio certificate issuance, artifact signing, Rekor transparency logging, and bundle verification. 
Our implementation extends the Sigstore signing path to collect additional scopes and embed \sysname proof material, then augments verification with policy checks over that proof. 
This strengthens identity assurance with minimal disruption to existing Sigstore workflows and no trusted hardware requirement, but does not protect against compromised signing clients.

\myparagraph{Trusted Authentication mode (\cref{fig:trusted_auth_mode})}
Attestation is enforced before signing by an external authority; in our prototype, this is realized through a local OIDC-mediated flow with a TEE-attested \sysname daemon. The daemon handles scope collection, proof generation, and CA interaction, reducing client-side trust assumptions. Sigstore remains in the trust path (including certificate issuance and Rekor logging), but attestation responsibility shifts from the client to the trusted CA. 
This improves resilience to compromised clients while retaining CA dependence.

\myparagraph{Core mode (\cref{fig:diverify_architecture})}
Attestation is verified directly by the verifier during \sysname proof evaluation, so CA-mediated attestation is no longer required in the attestation trust path. 
In our prototype flow, verification uses \sysname quote/proof checks directly rather than relying on the Sigstore transparency-and-certificate path as the primary attestation mechanism. 
This provides end-to-end enforcement of both identity and execution integrity, at the cost of requiring TEE-capable signing infrastructure.

\subsection{Implementation Details}
DiVerify implementation spans three elements: integrating scope providers, context-aware signing, and policy-based verification.

\myparagraph{Scope Providers}
We implement three scope providers covering the three claim categories from \cref{auth-mechanisms}. 
For identity claims, we use the GitHub OIDC token beacon~\cite{sigstoreConformanceBeacon} to obtain OIDC tokens from GitHub's identity provider. 
For possession claims, we implement a static provider that attests the signer's security key using YubiKey PIV attestation~\cite{yubico_piv_attestation}. 
For contextual claims, we implement a static provider that measures a device fingerprint. 
Static providers are wrapped using the strategy described in \S\ref{sec:multiscope} to bind their outputs to the signing session.

\myparagraph{Signing Client and Tool Attestation}
Because deployment modes place trust differently, we implement mode-specific signing paths. 
In Legacy-Compatible mode, we extend the Sigstore client to collect required claims from configured scope providers, construct a DiVerify proof, and include that proof in the standard Sigstore signing flow. 
In Trusted Authentication and Core modes, we split the client into an untrusted frontend and a trusted daemon. 
The frontend only accepts signing requests and forwards them to the daemon. The daemon runs inside an Intel SGX enclave using Gramine~\cite{gramine-image}, and performs scope retrieval, proof construction, key generation, and signing. Client advertises the set of scopes it supports using a trust level configuration.
Attestation quotes are produced via Intel SGX DCAP~\cite{IntelSGXPCCS}.

\myparagraph{Verifier and Policy Enforcement}
We also implement mode-specific verification paths. 
In Legacy-Compatible mode, we extend the Sigstore verifier to evaluate DiVerify policy over proof and signature, while retaining standard Sigstore checks (certificate and transparency-log verification). 
In Trusted Authentication mode, verification remains Sigstore-compatible, then applies the same policy evaluation. 
In Core mode, verification implementation is \sysname-native: the verifier checks quote-bound proof consistency and evaluates policy directly.
Attestation quotes are verified using Intel's quote verification library~\cite{IntelSGXDCAPQuoteVerification}.
Across all modes, policy evaluation logic is shared; only the source and enforcement point of trust evidence changes.
Policies are distributed as signed metadata and the verifier maintains local state recording the highest observed (\textsf{version}, \textsf{epoch}) pair per policy, enforcing monotonic policy evolution. 
We customized a local instance of Fulcio CA~\cite{sigstoreFulcio} to validate and embed DiVerify proof in issued certificates.

\myparagraph{Trust Levels Configuration}
\label{trust-level}
Our implementation exposes \emph{trust levels} that capture the minimum amount of signing-context information required for a valid signing event.
These levels are defined over the claim categories from~\cref{auth-mechanisms}, and are cumulative:

\begin{itemize}[leftmargin=2em,itemsep=2pt,topsep=2pt]
   \item \textit{Level 1 (Identity)}:
     An identity claim (\eg OIDC), providing baseline authentication.
   \item \textit{Level 2 (Identity + possession)}:
     Adds possession --- signers must control both identity and possession.
   \item \textit{Level 3 (Identity + possession + context)}:
     Further adds a contextual claim (\eg device/environment). 
\end{itemize}

\noindent
These levels are a coarse abstraction rather than a required part of the DiVerify model.
Other implementations could distinguish finer-grained boundaries,
\eg approved issuers or email domains within Level~1,
or
  stronger vs. weaker contextual evidence within Level~3,
or
  expose arbitrary policies.

\myparagraph{Implementation Cost}
To quantify the implementation effort, we report source lines of code (SLOC) as follows: Scope-provider integration, signing proof construction, and verifier policy evaluation add 1,620 LOC. SGX support requires TEE configuration (manifest and runtime), and Fulcio modifications to validate and embed proofs add 201 LOC.

\section{Formal Model, Soundness, and Security Analysis}\label{formal-model}
We formalize DiVerify verification as a predicate over the policy $P$, proof $\pi$, and artifact $A$. 
The desired security guarantee is this:
  \textit{if the verifier accepts $(P, \pi, A )$, then $\pi$ corresponds to a signing event satisfying the policy.}

\subsection{Syntax and Notation}

We use the following notation:
\begin{itemize}[leftmargin=1.4em,itemsep=2pt,topsep=2pt]
    \item \textbf{Proof.}
    $\pi = (u, sid, pk, \mathcal{C}, a, \sigma)$ bundles user $u$, session identifier $sid$, signing key $pk$, claims $\mathcal{C}$, attestation $a$, and signature $\sigma$.

    \item \textbf{Claim.}
    A claim $c_i \in \mathcal{C}$ has the form
    $c_i = (P_i, t, v, u_c, sid_c, \sigma_c)$,
    where provider $P_i$ signs $\sigma_c$ to assert that user $u_c$ has claim type $t$ with value $v$ in session $sid_c$.

    \item \textbf{Policy.}
    A policy is $P = (\mathsf{Scope}, \mathsf{Attest}, \mathsf{Rule})$,
    where $\mathsf{Scope}$ specifies relevant claim mappings,
    $\mathsf{Attest}$ specifies required environment evidence,
    and $\mathsf{Rule}$ is a Boolean condition over claims.

    \item \textbf{ValidClaim}$(c)$
    holds when the claim signature verifies under the provider.

    \item \textbf{ValidSig}$(\sigma, A, pk)$
    holds when $\sigma$ verifies under $pk$ as a signature over artifact $A$.

    \item \textbf{ValidAttest}$(a, sid, pk, M, Q)$
    holds when attestation $a$ verifies under procedure $Q$, reports measurement $M$, and is cryptographically bound to $(sid, pk)$.

    \item \textbf{Match}$(c, t, v)$
    holds when $ValidClaim(c)$ and $c$ matches expected claim type $t$ and value $v$.

    \item \textbf{Rule.}
    A rule has the grammar
    $R ::= \mathsf{Match}(c,t,v) \mid R_1 \wedge R_2 \mid R_1 \vee R_2$,
    \ie a Boolean expression over atomic predicates requiring specific claim types and values.

    \item \textbf{Witness}$(u, sid, \mathcal{C}, R, W)$
    mean that $W \subseteq \mathcal{C}$ is the set of claims witnessing satisfaction of rule $R$ for $u$ and session $sid$.

    \item \textbf{SameUser}$(W, u)$
    holds when all claims in $W$ used to satisfy the policy refer to the same user $u$.

    \item \textbf{SameSession}$(W, sid)$
    holds when all claims in $W$ used to satisfy the policy were issued for the same session $sid$.

    \item \textbf{DistinctProviders}$(W)$
holds when the claims in $W$ that witness the required rule
come from distinct providers.
\end{itemize}

\subsection{Lemmas and Theorem}
\label{sec:FormalModel:Theorem}

\subsubsection{Supporting Lemmas}

The lemmas below capture the properties underlying acceptance and reflect the requirements in~\cref{sec:multiscope,sec:attestation,sec:policy-semantics-and-lifecycle}.
Each follows from standard cryptographic guarantees or from the system assumptions in~\cref{sys-model}.

\begin{description}[leftmargin=0.5em,labelsep=0.4em,itemsep=3pt,topsep=2pt]
    \item[\textbf{Lemma 1 (Claim authenticity).}]
    If $\textsf{ValidClaim}(c)$, then $c$ was issued by its provider or the provider is compromised.
    This follows from provider signature unforgeability.

    \item[\textbf{Lemma 2 (Session binding).}]
    Assuming fresh session identifiers, $\textsf{SameSession}(W, sid)$ implies that all $c \in W$ originate from a single session.
    This relies on freshness and inclusion of $sid$ in all claims.

    \item[\textbf{Lemma 3 (Attestation binding).}]
    If $\textsf{ValidAttest}(a, sid, pk, M, Q)$, \\then the environment satisfies the specified measurement and binding, or the attestation mechanism is compromised.
    This follows from attestation authenticity and binding guarantees.
    
    \item[\textbf{Lemma 4 (Signer consistency).}]
    Assuming a canonical identity mapping across providers, $\textsf{SameUser}(W, u)$ implies that all $c \in W$ refer to the same user $u$.

    \item[\textbf{Lemma 5 (Witness validity).}]
    If $\mathsf{Witness}(u, sid, \mathcal{C}, R, W)$, then every atomic predicate in $R$ is supported by a claim in $W$ matching $u$ and $sid$.
    This follows by structural induction on $R$.

    \item[\textbf{Lemma 6 (Artifact binding).}]
    If $\textsf{ValidSig}(\sigma, A, sid, pk)$, then $A$ was signed under $pk$ in session $sid$, or the signing key is compromised.
    This follows from signature unforgeability.

    \item[\textbf{Lemma 7 (Provider threshold).}]
    If $\textsf{Accept}(P, \pi, A)$ and fewer than $k$ providers are compromised, then any policy requiring $k$ distinct providers includes at least one honest claim.
    This follows from Lemma~1 and the distinct-provider requirement.
\end{description}

\subsubsection{Theorem 1: Policy Enforcement Soundness}
\noindent

Let $P=(\textsf{Scope},\textsf{Attest},\textsf{Rule})$, 
$\pi=(u,sid,pk,\mathcal{C},a,\sigma)$, and artifact $A$. 
If $\textsf{Accept}(P,\pi,A)$, then there exists $W \subseteq \mathcal{C}$ such that:
(i) $W$ witnesses $\textsf{Rule}$, 
(ii) all $c \in W$ are valid and refer to $(u,sid)$, 
(iii) $\sigma$ binds $A$ to $(sid,pk)$, and 
(iv) if $\textsf{Attest}$ is required, the attestation corresponds to an environment satisfying a required measurement and trust boundary. 
$A$ was signed in a signing context satisfying $P$.


\myparagraph{Proof (Sketch)}
Assume
  $Accept(P,\pi,A)$,
  $\exists W\subseteq C$ such that ${SameUser}(W,u)$,
  that $ValidSig(\sigma,A,sid,pk)$, $Witness(u,sid,C,R,\\W)$,
  and
  SameSession(W,sid), $DistinctProviders(W)$.
Also, if required, suppose ${ValidAttest}(a, sid, pk, M, Q)$.


By Lemma 5, each predicate is supported by a matching claim in $W$, hence all witness claims are valid. By Lemmas 4 and 2, all claims refer to the same signer $u$ and session $sid$. By Lemma~6, $\sigma$ signs $A$ in session $sid$ under $pk$, except in compromise case. If attestation is required, then by Lemma 3, it satisfies the required measurement and binding properties.
Thus the accepted proof corresponds to a single signing event satisfying all policy requirements.

\myparagraph{Comments}
Acceptance requires a proof containing valid claims that form a witness $W$ for rule $R$, with all witness claims referring to the same user and session, and with the artifact signature (and attestation, if required) bound to that same context. Under the stated assumptions, violating this guarantee requires either compromising the relevant providers or the attestation mechanism, or breaking the authenticity or binding properties on which the proof relies.
\textit{\sysname raises the cost of a successful attack, rather than making policy violations impossible.}

\subsection{Security Analysis}
\begin{table}
    \caption{
    Compromise types by attack. $\Circle$ indicates a successful attack, where the verifier accepts a malicious signature. $\CIRCLE$ indicates prevention, where the verifier rejects it.
    }\label{tab:DiVerfy_comparison}
    \centering
    \small
    \begin{tabular}{lcccccc}
        \hline
        \bf Compromise & \bf Without & \multicolumn{3}{c}{\bf With DiVerify}\\ 
        \bf Type       & \bf DiVerify & \multicolumn{3}{c}{} \\
        \cline{3-5}
         &  & \bf Legacy & \bf Trusted & \bf Core \\ 
         &  & \bf Compat.& \bf Auth.   & \bf  \\
        \hline
        User Credential & $\Circle$ & $\CIRCLE$ & $\CIRCLE$ & $\CIRCLE$\\
        \hline 
        Scope Provider subset & $\Circle$  & $\CIRCLE$  & $\CIRCLE$  & $\CIRCLE$\\ 
        \hline
        Client & $\Circle$ & $\Circle$ & $\CIRCLE$ & $\CIRCLE$\\
        \hline
    \end{tabular}
\end{table}

We relate three attack scenarios (\cref{tab:DiVerfy_comparison}) in identity-based signing, derived from our threat model (\cref{sec:bg-threats}) and show how different DiVerify deployment variants (\cref{deploy-mode}) protect against them.

\paragraph{Types I \& II (Credential and Identity Provider Compromise)}
Under Type I, an attacker who compromises a signer's credentials can authenticate to a scope provider as that signer. Under Type II, a compromised or malicious scope provider can issue a claim for a signer who never authenticated. In both cases, the attack is confined to that scope provider.

The compromise is detected by the verifier during proof evaluation. By the threshold argument (Lemma 7), a proof containing only the one claim fails to form a complete witness set (Lemma 5), and the verifier rejects it.
 This check is identical across all three deployment modes, since scope evaluation is performed by the verifier in each mode.

The mitigation depends on the trust level (\cref{trust-level}) set in the policy. Level~1 requires only a single provider's claim, so the verifier's check above is satisfied by the attacker's one forged claim, and this compromise is not detected. A Level~2 or higher trust level policy requires the verifier to additionally match a claim from an independent provider, which the attacker does not control. This degrades the attack from successful to blocked as the required provider count increases.

\paragraph{Type III (Client Compromise)}

A compromised client may skip signer authentication and spoof scopes, producing a DiVerify proof that enables unauthorized signing. Whether this is caught depends on the operating mode, since attestation of the signing environment is produced outside the client's control only when an attester is enforced. In \emph{Legacy-Compatible} mode, no runtime attestation is enforced; a compromised client can use spoofed claims without an independent check, so this mode does not protect against Type III. In \emph{Trusted Authentication} mode, the CA validates the client integrity attestation before certificate issuance, so a compromised client cannot obtain a certificate without a valid attestation (Lemma~3). In \emph{Core} mode, the verifier checks attestation against the policy's expected attestation values at verification time. In both attested modes, a signature produced under a compromised environment fails attestation validation and is rejected.

\section{Quantitative Evaluation}
\label{sec:evaluation}

The formal model in~\cref{formal-model} establishes that DiVerify soundly enforces $S1 - S3$. 
This section evaluates the operational overhead that DiVerify introduces by answering the following questions:

\begin{itemize}[leftmargin=2em,itemsep=2pt,topsep=2pt]
    \item \textbf{Q1} What performance cost does DiVerify introduce?
    \item \textbf{Q2} What storage cost does DiVerify introduce? 
    \item \textbf{Q3} Can DiVerify integrate into existing ecosystems?
\end{itemize}

\myparagraph{Experiment Setup}
We ran all components in Docker containers on a 4.0 GHz Intel Xeon Platinum 8580 host with 499 GB RAM and SGX support, reflecting an industry-grade CI/CD environment~\cite{kalu2025industry}.
TEE components were built with Gramine~\cite{gramine-image}.
We deployed local Sigstore services (Fulcio, Rekor, and CTLog) and evaluated all three DiVerify deployment modes across trust levels 1--3, with 10 iterations per configuration. 

\subsection{Methodology}

\myparagraph{Q1--Performance cost}
To quantify the performance overhead of DiVerify, we measure end-to-end signing and verification latency across deployment modes and trust levels. For each run, we record the signing operations latency: scope collection, quote generation, proof construction, certificate issuance (Fulcio, when used), artifact signing, transparency-log submission (Rekor, when used). On verification, we measure policy verification and evaluation, certificate and identity checks, transparency-log/bundle verification (legacy and trusted auth. mode), quote/proof validation (Core mode), and final signature verification.
For each $(\text{mode}, \text{level})$ configuration, we present the mean measurement across iterations.

\myparagraph{Q2--Storage cost}
To quantify storage overhead, we measure the serialized verification artifact stored per run.
In Sigstore-style modes, this artifact is the Rekor bundle; in Core mode, it is the serialized \sysname proof and signature used for verification.

The signing certificate embeds the \sysname proof, so the main source of additional storage is proof metadata.
We use legacy authentication at trust level 1 as a baseline, since it relies only on standard OIDC-based  identity claims. 
We report overhead as the size difference between each other configuration and this baseline.

To assess practical impact, we sampled PyPI projects.
PyPI has $\sim$750,000 projects, each of which is associated with one or more released distribution files (packages).
We took a $\sim$1\% random sample, 10,000 projects, enumerating the associated packages (23,629, average size 1.8~MB).
We report overhead relative to package size.

\begin{table*}
    \caption{
    Average Signing \& Verification Overhead Introduced by DiVerify. 
    L1–L3 denote trust levels determining the required signer scopes.
    Attestation cost is paid by CA (Trusted Authentication), by verifier (Core), or not (Legacy-Compatible).
    }
    \label{table:sign_verify_cost}
    \centering
    \begin{tabular}{llccccccccc}
        \toprule
        \multicolumn{2}{l}{\textbf{}} & \multicolumn{3}{c}{\textbf{Legacy Compat. (ms)}} & \multicolumn{3}{c}{\textbf{Trusted Auth. (ms)}} & \multicolumn{3}{c}{\textbf{Core (ms)}} \\
        \cmidrule(lr){3-5} \cmidrule(lr){6-8} \cmidrule(lr){9-11}
        \multicolumn{2}{l}{\textbf{}} & L1 & L2 & L3 & L1 & L2 & L3 & L1 & L2 & L3 \\
        \midrule
        \multirow{3}{*}{\textbf{Signing}} 
        & Signing Time & 41.53 & 66.67 & 70.84 & 374.22 & 385.76 & 392.42 & 142.55 & 147.15 & 149.0 \\
        & Quote Gen.     & -- & -- & -- & 26.61 & 26.45 & 26.61 & 26.58 & 26.35 & 26.52 \\
        & Fulcio         & 29.63 & 30.05 & 30.80 & 242.98 & 242.85 & 242.75 & -- & -- & -- \\
        \midrule
        \multirow{3}{*}{\textbf{Verification}} 
        & Verify Time & 45.06 & 45.02 & 190.34 & 80.06 & 80.34 & 220.23 & 218.15 & 220.20 & 353.31 \\
        & Quote Verif.     & -- & -- & -- & -- & -- & -- & 205.54 & 207.22 & 206.74 \\
        \bottomrule
    \end{tabular}
\end{table*}

\subsection{Results}
\myparagraph{Q1--Performance cost}
DiVerify's overhead is not a fixed cost across deployments. 
As shown in \cref{table:sign_verify_cost}, Legacy-Compatible mode shows a low signing and verification latency, with Fulcio certificate issuance as the dominant contributor across trust levels. 
Verification cost is low for trust levels (\cref{trust-level}) L1 and L2, increasing significantly at L3 due to additional scope claim evaluation. With no attestation, this mode represents the minimum overhead DiVerify introduces.

Trusted Authentication mode incurs the highest signing latency, driven by the CA performing attestation validation before certificate issuance, which alone accounts for roughly 65\% of total signing time. 
This places the attestation burden at signing time, keeping verification comparatively low. 
Since Core mode removes the CA, it reduces signing latency by over 60\% relative to Trusted Authentication. 
However, since attestation shifts to the verifier, quote verification accounts for roughly 60\% or more of total verification time. 
The tradeoff is therefore not only when attestation cost is paid, but who bears the trust assumption: Trusted Authentication delegates attestation correctness to the CA transitively, while Core mode places it directly with the verifier.


\begin{table}[t]
\centering
\caption{
Storage overhead of DiVerify across modes.
Compared to baseline, DiVerify adds $\sim$0.5--28 KB, corresponding to $\sim$0.03--1.5\% of the average package size on PyPI. 
}
\begin{tabular}{lcrrr}
\toprule
\textbf{Config.} & \textbf{Level} & \textbf{Size (KB)} & \textbf{$\Delta$ (KB)} & \textbf{$\Delta$ / Pkg (\%)} \\
\midrule
\multirow{3}{*}{Legacy Compat.}
  & 1 & 11.2 & 0.0   & 0.00 \\
  & 2 & 11.7 & +0.5  & 0.03 \\
  & 3 & 16.1 & +4.8  & 0.26 \\
\midrule
\multirow{3}{*}{Trusted Auth.}
  & 1 & 34.9 & +23.7 & 1.28 \\
  & 2 & 35.4 & +24.2 & 1.31 \\
  & 3 & 39.7 & +28.5 & 1.54 \\
\midrule
\multirow{3}{*}{Core}
  & 1 & 7.1  & -4.1  & -0.22 \\
  & 2 & 7.2  & -4.0  & -0.22 \\
  & 3 & 8.4  & -2.9  & -0.15 \\
\bottomrule
\end{tabular}
\label{tab:storage-overhead}
\end{table}

\myparagraph{Q2--Storage cost}
As shown in~\cref{tab:storage-overhead}, for our Sigstore style signing systems (legacy compatible and trusted auth. mode) that log signatures, DiVerify adds between 0.5KB and 28.5KB per bundle depending on mode, with Trusted Authentication incurring the largest overhead due to the embedding of the DiVerify proof and attestation evidence into the issued certificate. \emph{Interestingly}, Core mode reduces bundle size below baseline by eliminating certificate issuance, though this reflects our current prototype's serialization format rather than a general lower bound. For package repositories, this overhead is negligible; even the worst case corresponds to under 1.6\% of the average PyPI package size, meaning the contextual provenance DiVerify adds is small relative to the artifact it protects.

\subsection{Case Studies}
\label{sec:case-studies}
To evaluate the feasibility and ecosystem compatibility of DiVerify (Q3), we conducted exploratory case studies through prototype integrations with three ecosystems.

\subsubsection{Method}
We selected three widely-used tools representing complementary roles in the software signing ecosystem: 
two signature producing systems (Sigstore and OpenPubkey) and one signature-consuming system (RSTUF~\cite{rstuf}). 
This selection allows us to evaluate whether DiVerify composes with both artifact generation and verification workflows.

For each case, we investigated: (1) the minimal integration effort required to achieve DiVerify’s security guarantees, and (2) how DiVerify composes with existing workflow artifacts and trust models. 

To achieve these, we reviewed publicly available deployments of each system to understand their workflows, and drafted integration documents describing how DiVerify could integrate. 
For OpenPubkey and RSTUF, we additionally engaged with project maintainers through virtual meetings to discuss and refine the proposed integration architecture.

\subsubsection{Results}

\myparagraph{Sigstore with DiVerify}
We integrated DiVerify with Sigstore following two models described above: legacy compatible and Trusted Auth.

For legacy, we extended the Sigstore signer to collect scopes and embed DiVerify proofs in the signing workflow and modified the verifier to perform policy evaluation against Sigstore's signature bundle.
The core Sigstore flow (OIDC - Fulcio - signing - Rekor submission) remains unchanged, requiring minimal pipeline changes. 
DiVerify proofs augment the signing certificate. 
Verification chains standard Sigstore validation (Rekor log entry checks, certificate validity) with policy-driven evaluation of DiVerify proofs, requiring minimal modifications to the verification flow.

Trusted Auth mode delegates scope collection and Fulcio interactions to an attested daemon, while still leveraging the Rekor transparency log, and the verifier only needs to handle daemon-generated signing certificates. 
In both modes, DiVerify proofs are embedded within issued certificates, allowing the verification logic to treat them as integral to the signature bundle.

We engaged with a maintainer from the Sigstore clients working group, who highlighted two key benefits: (1) DiVerify closes an existing gap in artifact verification by enforcing signing context, and (2) it enables expressing more assurance semantics aligned with organizational security requirements.

\myparagraph{OpenPubkey with DiVerify}
We implemented a DiVerify with OpenPubkey prototype by developing the OpenPubkey flow inside our codebase and then layering DiVerify attestation on top, rather than modifying an upstream OpenPubkey library. 

In legacy mode, the client collects required identity scopes through the OpenPubkey scope service, builds a DiVerify proof object (including OpenPubkey-derived identity claims and signer key material), and signs the artifact; verification then checks the PK token/CIC structure and signature chain, enforces DiVerify policy constraints, and validates the artifact signature. 
In core mode, scope handling and attestation are delegated to the attested daemon, which returns bundled signature material and DiVerify proof data for verification. 
The demo required implementing OpenPubkey core logic and integrating it to our scope collection, proof assembly, and verifier logic, with DiVerify wrapping the OpenPubkey token within the proof.

We then engaged an OpenPubkey maintainer to design a native integration that does not require wrapping the existing library. The resulting design identifies integration points and treats DiVerify proof as an artifact that wraps OpenPubKey token as an identity source, showing that \sysname can cleanly integrate with existing identity-based signing systems as a ``complex identity provider''.

This engagement surfaced two future directions. 
First, using GitHub Actions as an attestation oracle, with GitHub as the trust boundary providing verifiable organizational guarantees (as in~\cref{sec:attestation}), similar to URL Oracle~\cite{heilman_github_oracle_2025}. 
While suitable for CI workflows without additional hardware, this does not isolate the attester from the signing tool—the same gap addressed by TEE-based attestation in DiVerify’s core mode. 
Second, integrating additional scope providers to support heterogeneous identity systems across organizations.

\myparagraph{DiVerify with RSTUF}
\label{sec:rstuf-integration}
Integrating DiVerify with RSTUF allows package registries to enforce signing context requirements as a condition of package acceptance, extending the trust guarantees available to downstream verifiers.
This integration was carried out by using TUF-backed policy distribution as the trust bootstrap for DiVerify policies. 
DiVerify did not change RSTUF’s core repository workflow; instead, it consumes delegated policy targets from the RSTUF metadata/targets service using a TUF client update cycle (trusted root, metadata refresh, target resolution, and authenticated download). 
The retrieved policy is then evaluated against DiVerify proof material during signature verification. 
Integration effort is localized to the policy verification layer (\ie before metadata is presented to the TUF verifier). 
This shows that DiVerify can be layered onto an existing TUF trust model with minimal changes.


\section{Discussion}
\label{sec:discussion}


\subsection{Applying \sysname}
\label{sec:discussion-applying}

We consider how \sysname can be applied at three points in the software supply chain:
  upstream by artifact producers,
  at registries during publication and policy enforcement,
  and downstream by artifact consumers during verification.
\sysname complements mechanisms that secure adjacent stages of the lifecycle.

\myparagraph{(1) Upstream Integration}
\sysname can support upstream release governance by requiring build evidence, \eg provenance attestations such as SLSA~\cite{slsa_provenance}, as part of signing policy.
Under this model, an artifact is signed only if the signer can present both the required identity claims and evidence that the build satisfied the project's provenance requirements.
This would extend \sysname's scope-provider model to treat provenance services as contextual claim providers, turning build requirements from process conventions into machine-checkable signing conditions.

\myparagraph{(2) Registry Integration}
Registry-side use is a natural fit for \sysname.
Registries already mediate publication and enforce package-level policy.
For example, systems such as npm Trusted Publishing establish \emph{who} published a package, but not the signing conditions~\cite{npm-trusted-publishing}.
\sysname could strengthen this boundary by allowing a registry or package manager to verify both publisher identity and signing context before accepting or serving a package. 

\myparagraph{(3) Downstream Integration}
Downstream, \sysname enables consumers to express and enforce verifier-side trust policies rather than accepting any signature from an otherwise valid identity.
A client (\eg a deployment pipeline) could require that an artifact was signed under specific conditions before treating that artifact as acceptable.
This makes signing-context validation a first-class part of consumer-side verification and allows different adopters to tailor policies to their own risk models.

\vspace{0.20cm}
\textit{Is Policy a Barrier?}
Engineers may perceive \sysname's policy artifact as configuration overhead.
We assert that these policies only make explicit the decisions that maintainers already make implicitly, \eg when configuring trusted publishers on npm or PyPI~\cite{npm-trusted-publishing, pypi_trusted_publishers, pypi-digital-attestations}.
\sysname turns those manual choices into a structured policy artifact.

\subsection{\sysname and the End-to-End Principle}

More broadly, \sysname reflects an end-to-end design principle~\cite{saltzer1984end} for software signing: the property that matters is not that an intermediate service authenticated a signer, but that an artifact was signed under verifier-acceptable conditions. 
Intermediate components may contribute scopes, but validity must be enforced end-to-end by the verifier over the composed evidence. 
Hence, \sysname's policy layer is not configuration overhead; it provides a common mechanism for expressing and enforcing a trust contract between the upstream, any registry, and downstream. 
This perspective also justifies our deployment structure. 
Given that release infrastructure evolves across heterogeneous trust boundaries, practical signing mechanisms must support incremental deployment rather than assuming a single attestation model or uniform trust architecture.

\subsection{Limitations and Next Steps}
\label{sec:discussion-next}

\sysname can still be bypassed if an attacker can produce a signature that satisfies policy despite violating the verifier's real intent (\cref{sec:FormalModel:Theorem}).
What \sysname changes is the cost of doing so.
Stronger deployment modes require more specific and independent contextual claims, increasing the attacker's capabilities needs for a successful bypass. DiVerify's soundness guarantee holds relative to a given policy; it does not protect against a policy that is itself misconfigured or misaligned with the maintainer's actual threat model. Practices such as policy templates, deployment-specific defaults, and static validation for unsafe configurations can mitigate authoring risk, but do not eliminate the need for maintainers to define policies consistent with their threat model.

\sysname inherits a broader ecosystem limitation: its impact depends on the adoption and actual use of software signing.
Prior work shows that signing adoption remains uneven~\cite{schorlemmer2024signing}, that verification is often skipped in practice~\cite{kalu2025industry}, and that usability and integration problems continue to create friction even in identity-based ecosystems~\cite{kalu2025johnny,kalu2026longitudinal}.
Note that software signing remains an important target because cybersecurity frameworks, standards, and regulations emphasize software-supply-chain provenance~\cite{booth2025secure, chandramouli2024strategies}.

The case studies in~\cref{sec:case-studies} suggest that \sysname can compose with existing signing and distribution infrastructure without requiring those systems to be restructured.
Our next step is therefore to evaluate \sysname instantiations in realistic deployment settings.

\section{Conclusion}

Software signing has become central to supply chain security, with wide adoption of identity-based signing. At the same time, attackers have shifted toward the signing ecosystem components, targeting maintainer credentials, identity providers, and signing tools. However, existing verification approaches cannot enforce broader signing conditions and require implicit trust in the signing infrastructure. Prior work addresses parts of this space, from transparency logging, but does not enable machine checkable enforcement of how signing occurred.
DiVerify closes this gap by introducing multi identity verification, which prevents any single provider compromise from producing an accepted signature, environment attestation, which provides verifiable evidence that the signing was uncompromised, and policy governance, which makes acceptance conditions explicit and machine checkable, rather than implicit.
As signing becomes a compliance requirement across software supply chains, DiVerify enables release infrastructure to verify artifacts against explicit signing conditions instead of relying on implicit trust.

\bibliographystyle{ACM-Reference-Format}\balance
\bibliography{chinenye,bib}

\appendix
\end{document}